\documentclass[preprint,prl,10pt]{revtex4}%
\usepackage{amsfonts}
\usepackage{amsmath}
\usepackage{amssymb}
\usepackage{graphicx}%
\begin{document}
\preprint{cond-mat}
\title[ ]{Relaxation of the molecular state in atomic Fermi gases near a Feshbach resonance.}
\author{V. S. Babichenko}
\affiliation{Kurchatov Institute, Moscow.}
\pacs{PACS number}

\begin{abstract}
The relaxation processes in an ultra-cold degenerate atomic Fermi gas near a
Feshbach resonance are considered. It is shown that the relaxation rate of the
molecules in a resonance with the atomic Fermi system is of the order of the
fraction of the chemical potential determined by the interaction between
molecules. In this connection the lower part in the excitation spectrum of the
system of resonance molecules is not well-defined.

\end{abstract}
\volumeyear{ }
\volumenumber{ }
\issuenumber{ }
\eid{ }
\startpage{0}
\endpage{ }
\maketitle

The various investigations of atomic ultra-cold Fermi gases close to a
Feshbach resonance \cite{FFR} give a unique possibility to study many-body
problems both with the small and the strong couplings between particles. In
addition, the strength of the interaction can be regulated by changing the
magnitude of the external magnetic field \cite{FRE1} - \cite{FRE8}. There is a
lot of theoretical works in which the creation of the bound states of Fermi
atoms, i.e., molecules, in the degenerate Fermi gas near a Feshbach resonance
at ultra-low temperatures is considered. In majority of these works the
internal structure of molecules and the influence of the many-particle effects
on this structure are not taken into account \cite{HKCW}-\cite{AGR}. In these
works the molecules are considered as point particles and described by the
local boson field. However, in the case of the strong scattering of particles
the conditions for correctness of this supposition are violated and the
internal structure of molecules should be taken into account.

We consider the influence of the many-particle correlations on the creation of
bound states. In this connection, we take into account the internal structure
of molecules as well as the effect of the many-particle correlations on this
structure. The main attention is given to the case when one of the levels of
the bound state lies between zero and the Fermi energy of the atomic Fermi
gas. In this case some fraction of Fermi atoms, namely, those which have the
energy larger than the Feshbach resonance level, creates bound pairs with the
radius much smaller than the average spacing between atoms and these molecules
are in the resonance with the remaining Fermi gas. We obtain that, due to
many-particle correlations, the molecular propagator has the imaginary part.
This results in the relaxation of resonance molecules to the lower energy
state. The inverse relaxation time obtained is lager or about the fraction of
the Bose gas chemical potential connected with the interaction between
resonance molecules. \ 

The Hamiltonian of the model we consider is%

\begin{equation}
\widehat{H}=%
{\displaystyle\int}
d^{3}r%
{\displaystyle\sum\limits_{\substack{\alpha=1,2;\\\sigma=\uparrow,\downarrow
}}}
\widehat{\Psi}_{\alpha,\sigma}^{+}\left(  \overrightarrow{r}\right)
\varepsilon_{\sigma}\left(  \widehat{\overrightarrow{p}}\right)  \widehat
{\Psi}_{\alpha,\sigma}\left(  \overrightarrow{r}\right)  +\widehat{H}%
_{int}+\widehat{H}_{tr} \tag{1}%
\end{equation}

where%

\begin{equation}
\widehat{H}_{int}=%
{\displaystyle\int}
d^{3}rd^{3}r^{\prime}%
{\displaystyle\sum\limits_{\substack{\sigma_{1};\sigma_{2}\\\alpha_{1}%
,\alpha_{2}}}}
g_{\sigma_{1},\sigma_{2}}\left(  \overrightarrow{r}-\overrightarrow{r}%
^{\prime}\right)  \widehat{\Psi}_{\alpha_{1},\sigma_{1}}^{+}\left(
\overrightarrow{r}\right)  \widehat{\Psi}_{\alpha_{2},\sigma_{2}}^{+}\left(
\overrightarrow{r}^{\prime}\right)  \widehat{\Psi}_{\alpha_{2},\sigma_{2}%
}\left(  \overrightarrow{r}^{\prime}\right)  \widehat{\Psi}_{\alpha_{1}%
,\sigma_{1}}\left(  \overrightarrow{r}\right)  \tag{1.a}%
\end{equation}

\[
\widehat{H}_{tr}=t%
{\displaystyle\int}
d^{3}r\left(
{\displaystyle\sum\limits_{\alpha_{1}\neq\alpha_{2}}}
\widehat{\Psi}_{\alpha_{1},\uparrow}^{+}\left(  \overrightarrow{r}\right)
\widehat{\Psi}_{\alpha_{2},\downarrow}^{+}\left(  \overrightarrow{r}^{\prime
}\right)  \widehat{\Psi}_{\alpha_{2},\uparrow}\left(  \overrightarrow
{r}^{\prime}\right)  \widehat{\Psi}_{\alpha_{1},\uparrow}\left(
\overrightarrow{r}\right)  +h.c.\right)
\]

and%

\[
\varepsilon_{\uparrow}\left(  \overrightarrow{p}\right)  =\frac
{\overrightarrow{p}^{2}}{2m}-\mu_{F};\text{ \ \ }\varepsilon_{\downarrow
}\left(  \overrightarrow{p}\right)  =\frac{\overrightarrow{p}^{2}}{2m}%
+E_{0}-\mu_{F};\text{ \ \ \ }g_{\uparrow,\downarrow}<0
\]

The operator $\widehat{\overrightarrow{p}}=-i\overrightarrow{\nabla}$ \ is the
momentum operator. The arrow indexes $\sigma=\uparrow,\downarrow$ of Fermi
operators $\widehat{\Psi}_{\alpha,\sigma}$ denote the direction of the
electron spin of atoms with respect to the external magnetic field. The
direction $\uparrow$ of the atomic electron spin means the orientation of the
spin along the direction of the external magnetic field, indexes $\alpha$
denote the direction of the nuclear spin. The interaction between electron and
nuclear spins is treated as a small value. Energy $E_{0}$ is determined by the
interaction between the external magnetic field $h$, which is supposed to be
homogeneous, and the atomic spin. It has the magnitude $E_{0}=2\mu_{e}h$ where
$\mu_{e}$ is the Bohr magneton.\ The term $\widehat{H}_{tr}$ in the
Hamiltonian is connected with the possibility of hybridization between the
singlet closed channel and the triplet open channel. In this paper we neglect
hybridization $\widehat{H}_{tr}$ and suppose that conditions for the Feshbach
resonance, which are formulated below, are fulfilled. Note that the
interaction between atoms $g_{\sigma_{1},\sigma_{2}}\left(  \overrightarrow
{r}-\overrightarrow{r}^{\prime}\right)  $ is independent of nuclear spins
$\alpha$ and depends on the relative orientations of the atomic electron spins
of interacting atoms $\sigma=\uparrow,\downarrow$. The value $g_{\uparrow
,\downarrow}$ is supposed to correspond to a strong attractive potential. At
the same time, interaction potentials $g_{\uparrow,\uparrow\text{ }}$,
$g_{\downarrow,\downarrow\text{ }}$are assumed to have a small value or being
repulsive ones.

The propagator of bound states of atoms, i.e. molecules, is determined by a
sum of the ladder diagrams which can be found by solving the Bethe-Salpeter
equation. This equation can be written in the form%

\begin{equation}
T=T^{\left(  0\right)  }+T^{\left(  0\right)  }g_{\uparrow,\downarrow}T
\tag{2}%
\end{equation}

Here $T^{\left(  0\right)  }$ is one link of the ladder diagram in which the
integration over the internal four-momentum is performed over zero component
of this momentum, i.e. over the internal frequency alone. It has the form%

\begin{equation}
T^{\left(  0\right)  }\left(  P,p,p^{\prime}\right)  =-i%
{\displaystyle\int}
\frac{d\varepsilon}{2\pi}G_{\uparrow}^{\left(  0\right)  }\left(  \frac{P}%
{2}+p\right)  G_{\downarrow}^{\left(  0\right)  }\left(  \frac{P}{2}-p\right)
\delta_{pp^{\prime}} \tag{3}%
\end{equation}

The momentums $P$ and $p$ are the total four-momentum and the four-momentum of
the relative motion of two particles, respectively, and $\varepsilon$ is zero
component or the frequency of the relative motion. Note that the integration
over the internal frequency $\varepsilon$ can be performed in each link of the
ladder diagrams since the interaction is time-independent.\ The calculation of
$T^{\left(  0\right)  }$ gives%

\begin{equation}
T^{\left(  0\right)  }\left(  P,p,p^{\prime}\right)  =\frac{1}{\Omega
-\frac{\overrightarrow{P}^{2}}{4}-\overrightarrow{p}^{2}-E_{0}+2\mu
_{F}+i\delta}\delta_{pp^{\prime}} \tag{4}%
\end{equation}

Here and henceforth we choose the system of units so as the magnitude of the
atomic mass is equal to unity m=1. Formally, the solution of Eq.(2) can be
rewritten in the form%

\begin{equation}
T=\frac{1}{\left(  T^{\left(  0\right)  }\right)  ^{-1}-g_{\uparrow
,\downarrow}} \tag{5}%
\end{equation}

It can easily be seen that Eq.(5) is analogous to the expression for the Green
function of one particle with the reduced mass in the external field
$g_{\uparrow,\downarrow}\left(  \overrightarrow{r}\right)  $. The solution of
this equation can be written in the form%

\begin{equation}
T\left(  P,p,p^{\prime}\right)  =%
{\displaystyle\sum\limits_{n}}
\frac{\psi_{n}\left(  \overrightarrow{p}\right)  \overline{\psi}_{n}\left(
\overrightarrow{p}^{\prime}\right)  }{\Omega+2\mu_{F}-\frac{\overrightarrow
{P}^{2}}{4}-E_{0}-E_{n}+i\delta}+%
{\displaystyle\int}
\frac{d^{3}k}{\left(  2\pi\right)  ^{3}}\frac{\psi_{_{\overrightarrow{k}}%
}\left(  \overrightarrow{p}\right)  \overline{\psi}_{_{\overrightarrow{k}}%
}\left(  \overrightarrow{p}^{\prime}\right)  }{\Omega+2\mu_{F}-\frac
{\overrightarrow{P}^{2}}{4}-E_{0}-\overrightarrow{k}^{2}+i\delta} \tag{6}%
\end{equation}

Wave functions $\psi_{n}$, $\psi_{_{\overrightarrow{k}}}$ are the
eigen-functions of Hamiltonian $\widehat{H}_{m}=\widehat{\overrightarrow{p}%
}^{2}+g_{\uparrow,\downarrow}\left(  \overrightarrow{r}\right)  $. The
functions $\psi_{n}$, $\psi_{_{\overrightarrow{k}}}$\ correspond to the
discrete and continuous spectra, respectively, and obey the equations%

\begin{equation}
\widehat{H}_{m}\psi_{n}=E_{n}\psi_{n}\text{; \ \ \ \ \ \ \ }\widehat{H}%
_{m}\psi_{\overrightarrow{k}}=\overrightarrow{k}^{2}\psi_{\overrightarrow{k}}
\tag{7}%
\end{equation}

In the case of the Feshbach resonance there is a discrete level $E_{n_{0}%
}\equiv E_{Fesh}\approx-E_{0}$. Usually, the conditions of experiment are such
that this discrete level is not the lowest one. The existence of the Feshbach
resonance means that the energy $\Delta E=E_{0}-E_{Fesh}$ obeys inequalities
$\mid\Delta E\mid<<E_{0}$ and $\mid\Delta E\mid\lesssim\varepsilon_{F}$.
Different states $\psi_{n}$ of the discrete spectrum $E_{n}$ describe the
different states of molecules and the propagators of molecules in these states
have the form of the terms in the discrete sum of Eq.(6). At the same time,
the second summand in Eq.(6) determines the molecular states of the continuous
part of the molecular spectrum.

Considering the ladder diagrams alone, we take the two-particle processes into
account and neglect the many-particle correlations completely. To involve
these correlations, we dress the fermion Green functions in the ladder
diagrams by the self-energy part. In the approximation of the point-like
interaction between atoms $g\left(  \overrightarrow{r}\right)  =g_{aa}%
\delta\left(  \overrightarrow{r}\right)  $ this self-energy part can be
written in the form%

\begin{equation}
\Sigma_{F}\left(  x_{1},x_{2}\right)  =ig_{aa}^{2}G\left(  x_{1},x_{2}\right)
G\left(  x_{1},x_{2}\right)  G\left(  x_{2},x_{1}\right)  \tag{8}%
\end{equation}

where $g_{aa}$ is the effective interaction between atoms. The self-energy
part $\Sigma_{F}$ has an imaginary part. Its retarded component for the large
external momentum $\overrightarrow{p}$, compared with the Fermi momentum
$p_{F}$\ of the atomic gas, can be calculated as%

\begin{equation}
\operatorname{Im}\Sigma_{F\uparrow}^{\left(  R\right)  }\left(  p\right)
=\gamma=-g_{aa}^{2}\mid\overrightarrow{p}\mid\rho\tag{9}%
\end{equation}

where $\rho$ is the density of the Fermi gas. Note that, if we consider a
portion in a sum of the ladder diagrams corresponding to the molecule
propagator, the average value $\mid\overrightarrow{p}\mid_{av}$ of the
momentum $\mid\overrightarrow{p}\mid$ in this case is of the order of the
inverse radius of the bound state. In the case of the small density of Fermi
gas it obeys inequality $\mid\overrightarrow{p}\mid_{av}>>p_{F}$. Due to
finite imaginary part in the self-energy part of Fermi propagator, the finite
imaginary part appears also in the propagator of a molecule, i.e., in the
terms of a discrete sum in Eq. (6). The calculation of a sum of the ladder
diagrams with the dressed fermion propagators under conditions of the thermal
equilibrium of the Fermi gas in the Keldysh - Schwinger technique \cite{Schw},
\cite{Kel} gives the following expressions for the retarded $D^{\left(
R\right)  }$, advanced $D^{\left(  A\right)  }$, and kinetic $D^{\left(
K\right)  }$ propagators of molecules%

\begin{equation}
D^{\left(  R,A\right)  }\left(  \Omega\right)  =\frac{1}{\Omega+2\mu_{F}%
-\frac{\overrightarrow{P}^{2}}{4}-E_{0}-E_{Fesh}\pm i\Gamma_{1}} \tag{10}%
\end{equation}

\begin{equation}
D^{\left(  K\right)  }\left(  \Omega\right)  =\coth\left(  \frac{\Omega}%
{2T}\right)  \left[  D^{\left(  R\right)  }\left(  \Omega\right)  -D^{\left(
A\right)  }\left(  \Omega\right)  \right]  \tag{11}%
\end{equation}

The value $E_{Fesh}<0$ is the binding energy of the molecule. It is necessary
to emphasize that Eq.(11) is not an assumption which can be made on the
natural basis when the Fermi gas state is supposed to be equilibrium one. This
can be obtained in the Keldysh - Schwinger technique by direct summation of
the ladder diagrams with the bare link (3) consisting of the equilibrium Fermi
propagators. Note that, in the case of $\Delta E=E_{0}-\mid E_{Fesh}\mid<0$
near the pole of $D^{\left(  R,A\right)  }$, the value of $D^{\left(
K\right)  }\left(  \Omega,\overrightarrow{P}\right)  $ becomes negative in the
case $\Gamma_{1}>0$. However, this fact is in contradiction with the general
properties of $D^{\left(  K\right)  }\left(  \Omega,\overrightarrow{P}\right)
$ \cite{Schw}. The change of Green functions $D^{\left(  R\right)
}\rightarrow D^{\left(  A\right)  }$, and vice versa, in Eq.(11) with the
negative value of $\coth\left(  \frac{\Omega}{2T}\right)  $ restores the
correct sign of $D^{\left(  K\right)  }$. This change in fact means a change
of the sign of $\Gamma_{1}$. Thus, in the case of $\Delta E<0$ near the pole
of $D^{\left(  R,A\right)  }$\ the value $\Gamma_{1}$ becomes negative
$\Gamma_{1}<0$. This, in its turn, results in an instability of the system and
exponential growth of the number of molecules. At the same time, the Fermi
distribution becomes unstable too.

The imaginary part $\Gamma_{1}$ of the self-energy part of the molecular
propagator can be estimated just as value (9) with the momentum $p$ of about
inverse radius of a molecule, i.e., $p\sim\sqrt{\mid E_{Fesh}\mid}$,%

\begin{equation}
\Gamma_{1}\sim g_{aa}^{2}\sqrt{\mid E_{Fesh}\mid}\rho\sim\frac{1}{\sqrt{\mid
E_{Fesh}\mid}}\rho\tag{12}%
\end{equation}
\ 

The relaxation $\Gamma_{1}$\ near the pole of propagator $D^{\left(
R,A\right)  }$\ is practically independent of frequency $\Omega$ and momentum
$\overrightarrow{P}$ for small values of $\Omega$ and $\overrightarrow{P}$
compared with the bounding energy of a molecule and its inverse radius,
simultaneously. This fact distinguishes essentially the Bose gas of resonance
molecules coexisting with the degenerate Fermi gas of atoms from the Bose gas
of molecules in the lack of degenerate Fermi gas of atoms.\ The physical
reason for an existence of this imaginary part is due to creation of Fermi gas
excitations as a result of the interaction of the Fermi gas with the molecule.
Note that the existence of the Cooper gap at the Fermi surface has no
influence on the magnitude $\Gamma_{1}$ because the main contribution to the
link of ladder diagrams is given by the large momentums much larger than the
Fermi momentum $p_{F}.$

The diagrams more complicated than considered before, for the self-energy part
of the molecular propagator, namely, the diagrams involving intersections of
the interaction lines with the ladder links can be taken into account. These
diagrams take into account the internal structure of molecules and describe
the process in which, due to the atom-molecule interaction, the molecule
transfers to the lower energy state, strongly exciting an atom and obtaining
the momentum almost opposite to the momentum of the excited atom. This process
results in the relaxation of the resonance molecular system to the lower
energy state. The self-energy part corresponding to this diagram has the form%

\begin{equation}
\Sigma_{M}^{\left(  n\right)  }\left(  x_{1},x_{2}\right)  =ig_{aM}^{2}%
D_{M}^{\left(  n-1\right)  }\left(  x_{1},x_{2}\right)  G_{F}\left(
x_{1},x_{2}\right)  G_{F}\left(  x_{2},x_{1}\right)  \tag{13}%
\end{equation}

Here $g_{aM}$ is the effective interaction between atom and resonance
molecule, $D_{m}^{\left(  n-1\right)  }\left(  x_{1},x_{2}\right)  $ denotes
the propagator of the molecule which has the energy $E_{n-1}$ lower than the
energy $E_{n}$ corresponding to the molecular self-energy part $\Sigma
_{M}^{\left(  n\right)  }\left(  x_{1},x_{2}\right)  $. Note that the
atom-resonance molecule effective interaction $g_{aM}$ is significantly larger
than the atom-atom interaction $g_{aa}$ due to large size of resonance
molecule compared with the size of atoms. The imaginary part of this
self-energy part can be written as%

\[
\operatorname{Im}\Sigma_{M}^{\left(  n\right)  }\left(  \Omega,\overrightarrow
{P}\right)  =\rho g_{am}^{2}%
{\displaystyle\int}
\frac{d^{3}p}{\left(  2\pi\right)  ^{3}}\delta\left[  \left(  \Omega
-\frac{p^{2}}{4}-E_{n-1}\right)  -\left(  \frac{\left(  -\overrightarrow
{p}+\overrightarrow{P}\right)  ^{2}}{2}-\frac{p_{F}^{2}}{2}\right)  \right]
\]

Thus, the relaxation rate of the molecular state $\Gamma_{2}=\operatorname{Im}%
\Sigma_{m}^{\left(  n\right)  }$ can be calculated as%

\[
\Gamma_{2}\sim\rho g_{aM}^{2}\sqrt{\mid E_{n-1}\mid}\sim\rho g_{aM}^{2}%
\sqrt{\mid E_{aM}^{\left(  bound\right)  }\mid}\sim\frac{\rho}{\sqrt{\mid
E_{aM}^{\left(  bound\right)  }\mid}}%
\]

In Eq. (12) $\Sigma_{M}^{\left(  n\right)  }$ denotes the self-energy part of
the molecule which corresponds to such molecular binding energy that the
energy of the molecule $E_{n}$ in the centre-of-mass frame of the molecule is
in the resonance with the atomic Fermi gas, i.e., $\Sigma_{M}^{\left(
n\right)  }$ is the self-energy part of resonance molecules. At the same time,
the molecular propagator $D_{M}^{\left(  n-1\right)  }$ corresponds to the
lower bound state of the molecule. The energy of such molecule in the case of
zero total momentum has value $E_{n-1}<0$. The energy $E_{aM}^{\left(
bound\right)  }$ is the binding energy of a resonance molecule and an atom,
and $E_{n-1}=E_{n}-\mid E_{aM}^{\left(  bound\right)  }\mid$.

If the bare value of the molecular level is less than the Fermi level of the
atomic system and positive, a fraction of fermions bound to the molecules,
such that the molecular level lies at the Fermi level of the residuary Fermi
system. During the times of about $1/\Gamma_{2}$\ the created resonance
molecules transfer to the lower energy state and escape from the resonance.
Note that the value of this relaxation rate is of the order of the chemical
potential in the system of resonance molecules determined by the interaction
$g_{MM}\sim g_{aM}$ between molecules. As the result, the lower part of the
spectrum in the system of resonance molecules has a large imaginary part and
is not well-defined.

This work was supported by the RFFI grant.


\begin{thebibliography}{99}                                                                                               %


\bibitem {FFR}W.C. Stwalley, Phys. Rev. Lett. 37, 1628 (1976); E. Tiesinga,
B.J. Verhaar, and H.T.J. Stoof, \ Phys. Rev. A 47, 4114 (1993); S. Inouye,
M.R. Andrews, J. Stenger, H.-J. Miesner, D.M. Stamper-Kurn, and W. Ketterle,
Nature (London) 392,151 (1998).

\bibitem {FRE1}M. Greiner, C.A. Regal, and D.S. Jin, Nature (London), 426, 537
(2003). \ 

\bibitem {FRE2}S. Jochim, M.Bartenstein, \ A. Almeyer, G. Hendl, S. Reidl, C.
Chin, J.H. Denschlang, and R. Grimm, Science, 302, 2101 (2003). \ 

\bibitem {FRE3}M.W. Zwierlein, C.A. Stan, C.H. Schunck, S.M.F. Raupach, S.
Gupta, Z. Hadzibabic, and Ketterle, Phys. Rev. Lett., 91, 250401, (2003).

\bibitem {FRE4}T. Bourdel, L. Khaykovich, J. Cubizolles, Z. Zhang, F. Chevy,
M. Teichmann, L. Tarruell, S.J.J.M.F. Kokkelmans, and C. Salomon, Phys. Rev.
Lett., 93, 050401 (2004).

\bibitem {FRE5}C.A. Regal, M. Greiner, and D.S. Jin, Phys. Rev. Lett. 92,
040403 (2004).

\bibitem {FRE6}M.W. Zwierlein, C.A. Stan, C.H. Schunck, S.M.F. Raupach, A.J.
Kerman, and W. Ketterle, Phys. Rev. Lett., 92, 120403 (2004).

\bibitem {FRE7}J. Kinast, S.L. Hemmer, M.B.Gehm, A. Turlapov, and J.B. Thomas,
Phys. Rev. Lett., 92, 150402 (2004).

\bibitem {FRE8}M.W. Zwierlein, C. H. Schunck, C. A. Stan, S.M.F. Raupach, and
W. Ketterle, Phys. Rev. Lett., 94, 180401 (2005).

\bibitem {HKCW}M. Holland, S.J.J.M.F. Kokkelmans, M.L. Chiofalo, and R.
Walser, Phys.Rev.Lett. 87, 120406 (2001).

\bibitem {OG}Y. Ohashi and A. Griffin, Phys. Rev. A, 67, 063612 (2003).

\bibitem {FH}G.M. Falco and H.T.C. Stoof, Phys. Rev. Lett., 92, 130401 (2004).

\bibitem {KSS}S.J.J.M.F. Kokkelmans, G.V. Shlyapnikov, C. Salomon, Phys. Rev.
A, 69, 031602 (2004).

\bibitem {BL}R.A. Barankov and L.S. Levitov, Phys.Rev.Lett. 93, 130403 (2004).

\bibitem {BLS}R.A. Barankov, L.S. Levitov, and B.Z. Spivak, Phys.Rev.Lett. 93,
160401 (2004).

\bibitem {AGR}A.V. Andreev, V. Gurarie, L. Radzihovsky, Phys.Rev.Lett. 93,
130402 (2004)

\bibitem {Schw}J. Schwinger, J. Math. Phys. 2, 407 (1961).

\bibitem {Kel}L.V. Keldysh, JETP, 47, 1515, (1964).
\end{thebibliography}
\end{document}